\documentclass[paper]{JHEP3} 

\JHEPspecialurl{http://jhep.sissa.it/JOURNAL/JHEP3.tar.gz}

\usepackage{epsfig,multicol}
\usepackage{cite}

\usepackage{amsmath}
\usepackage{amssymb}
\preprint{HU-EP-04/20\\
KEK-TH-952\\
hep-th/0404179\\}

\title{
On the relation between
non-commutative field theories at $\theta =\infty$ 
and large $N$ matrix field theories
}

\author{Wolfgang Bietenholz and Frank Hofheinz \\
        Institut f\"{u}r Physik, Humboldt Universit\"{a}t zu Berlin \\
        Newtonstr. 15, D-12489 Berlin, Germany \\
        E-mail: \email{bietenho@physik.hu-berlin.de, 
        hofheinz@physik.hu-berlin.de}}

\author{Jun Nishimura \\
        High Energy Accelerator Research Organization (KEK)\\
        1-1 Oho, Tsukuba 305-0801, Japan \\
        E-mail: \email{jnishi@post.kek.jp}}

\abstract{
It is well-known that 
non-commutative (NC) field theories at $\theta =\infty$ 
are ``equivalent'' to large $N$ matrix field theories
to all orders in perturbation theory,
due to the dominance of planar diagrams.
By formulating a NC field theory on the lattice
non-perturbatively and mapping it onto a twisted reduced model,
we point out that the above equivalence does not hold if the translational
symmetry of the NC field theory is broken spontaneously.
As an example we discuss NC scalar field theory, where
such a spontaneous symmetry breakdown  
has been confirmed by Monte Carlo simulations.
}


\newcommand{\id}{{1\!\!1}} 
\newcommand{\bel}{\begin{equation}\label}

\newcommand{\non}{\nonumber \\}

\newcommand {\beq}{\begin{equation}}
\newcommand {\eeq}{\end{equation}}
\newcommand {\beqa}{\begin{eqnarray}}
\newcommand {\eeqa}{\end{eqnarray}}
\newcommand {\bc}{\begin{center}}
\newcommand {\ec}{\end{center}}

\newcommand {\Tr}{\mbox{Tr\,}}

\newcommand {\ee}{\mbox{e}}

\def\dag{\dagger}

\def\vs5{\vspace*{5mm}}
\def\vs1{\vspace*{1cm}}
\def\vs2{\vspace*{2cm}}
\def\hs5{\vspace*{5mm}}
\def\hs1{\hspace*{1cm}}
\def\hs2{\hspace*{2cm}}
\def\vs50{\vspace*{50mm}}
\def\vs20{\vspace*{20mm}}

\begin{document}


\section{Introduction}

Field theories on non-commutative (NC) spaces,
which are characterized by the commutation relation
among the coordinate operators $\hat{x}_\mu$ ($\mu = 1 , \cdots , D$),
\beq
[ \hat{x}_\mu , \hat{x}_\nu  ] = i \, \Theta_{\mu\nu } \  ,
\label{NC_rel}
\eeq
have recently attracted
much attention in the context of string theory and quantum gravity.
In particular perturbative aspects of such theories have been
discussed extensively in the literature. Diagrammatically the difference from
ordinary field theories in the commutative space is represented
by a momentum dependent phase factor of the form 
\beq
\exp \left( - \frac{i}{2} \,  \Theta_{\mu \nu}
\sum_{i<j} p^{(i)}_\mu  p^{(j)}_\nu \right)
\eeq
assigned to each vertex, 
where $\Theta_{\mu\nu}$ is the non-commutativity tensor
in eq.\ (\ref{NC_rel}),
and $p^{(i)}_\mu$ ($i = 1,\cdots , n $) represent
the momenta flowing into the vertex.
The phase factors (except those depending
only on the external momenta) cancel in the planar diagrams,
whereas in the non-planar diagrams, at least one phase factor 
depending on an internal momentum remains.

Let $\theta$ be the typical magnitude of the non-zero elements of the
non-commutativity tensor $\Theta_{\mu\nu}$.
If we take the $\theta \rightarrow \infty$ limit
(after removing the phase factor depending only on the external momenta),
the non-planar diagrams vanish due to the oscillating phase,
and we are left with the planar diagrams.
In this way the $\theta \rightarrow \infty$ limit of NC field theories
is ``equivalent'' to the large $N$ matrix field theory in the
{\em commutative} space to all orders in perturbation theory.

In this letter we point out that this well-known ``equivalence''
does not always hold {\em non-perturbatively}.
In order to discuss this issue, we definitely
need a non-perturbative formulation of NC field theory. 
As in ordinary field theories, we may regularize a NC field theory
non-perturbatively by putting it on a lattice \cite{AMNS}.
This is achieved by imposing the operator identity
\beq
\ee ^{2 \, \pi \, i \, \hat{x}_\mu / a } = \id \ ,
\label{UVreg}
\eeq
where $a$ is the lattice spacing,
and the lattice sites are ``fuzzy'' due to the non-commutativity 
(\ref{NC_rel}).
Using a correspondence between a field on a finite lattice and a matrix,
which is {\em exact}
in the sense that the {\em finite} degrees of freedom on one side are mapped
onto the other, one to one,
we can map a lattice NC field theory
onto the twisted reduced model \cite{AMNS}. 
%
%
%
%
%
This is a refinement of the earlier work
\cite{Aoki:1999vr},
where continuum NC theory is mapped to the 
continuum version of the twisted reduced model \cite{Gonzalez-Arroyo:1983ac},
which is formally defined in terms of infinite-dimensional matrices.

Reduced models appeared in history first as an equivalent 
description of large $N$ gauge theory in the early 80s 
\cite{Eguchi:1982nm}.
The original Eguchi-Kawai (EK) model is proved to be ``equivalent'' to 
SU($\infty$) lattice gauge theory
under the assumption that the U(1)$^D$ symmetry
of the model is {\em not} spontaneously broken.
It was found, however, that this symmetry {\em is}
actually broken in the weak coupling regime 
at $D\ge 3$ \cite{Bhanot:1982sh}.
The {\em twisted} EK model is one of the proposals
to overcome this problem \cite{Gonzalez-Arroyo:1982hz}
\footnote{For other proposals, which may be suitable for
numerical studies of planar QCD, see Refs.\ \cite{NN}.},
and the equivalence was confirmed
by the weak coupling expansion 
as well as the strong coupling expansion.
Soon later it was noticed that this idea of ``twisted reduction''
can be applied also to non-gauge theories written in terms of matrix fields
\cite{Eguchi:1982ta,Gonzalez-Arroyo:vx}.
As in the case of gauge theory 
\cite{Eguchi:1982nm,Gonzalez-Arroyo:1982hz},
the proof based on the Schwinger-Dyson (SD) equations
reveals that there are certain symmetries
which should not be spontaneously broken 
in order for the equivalence to hold \cite{Gonzalez-Arroyo:vx}.

In the new interpretation of the twisted reduced model as
a lattice regularization of NC field theories,
the non-commutativity parameter $\theta$ 
is identified as \cite{AMNS}
\beq
\theta  
\propto
 N^{2/D} a^2 \ ,
\label{theta_identify}
\eeq
where $N$ is the size of the matrices that appear in
twisted reduced models.
In the context of EK equivalence, one has to take
the large $N$ limit at a fixed lattice spacing $a$
(hence at fixed bare parameters with appropriate normalization with
respect to $N$)
\footnote{This large $N$ limit is referred to as the 
{\em planar} large $N$ limit,
since in perturbation theory only planar diagrams survive this limit.
On the other hand, in order to obtain the continuum limit 
of NC field theories with finite $\theta$, 
one has to take the large $N$ limit
and the $a \rightarrow 0$ limit simultaneously,
which is usually referred to as the {\em double scaling} limit.
Indeed various correlation functions scale in this limit
for simple models as demonstrated by Monte Carlo simulations
\cite{2dU1,Procs,BHN}.},
which corresponds to the $\theta =\infty$ limit of NC field theories
as is clear from eq.\ (\ref{theta_identify}).
Thus the EK equivalence provides a non-perturbative account for the
aforementioned ``equivalence''
between the NC field theories at $\theta =\infty$ and 
the large $N$ matrix field theories. 
We will see that the condition for this ``equivalence'' is
--- in the language of NC field theories ---
that the translational symmetry is not
spontaneously broken.

In fact in NC scalar field theories 
the spontaneous breakdown of translational invariance
was conjectured in Refs.\ \cite{Gubser:2000cd,CZ},
and confirmed recently by Monte Carlo simulations 
in Refs.\ \cite{Procs,Ambjorn:2002nj,BHN}.
Before we discuss the NC scalar field theory, we start
in the next Section with a brief comment on the gauge theory case.
For a comprehensive review on the EK equivalence 
we refer the reader to Ref.\ \cite{Das:1984nb},
and for reviews on 
the novel connection between reduced models and NC field theories
to Refs.\ \cite{Szabo:2001kg,Nishimura:2003rj}.

\section{The case of gauge theory}

The twisted EK model \cite{Gonzalez-Arroyo:1982hz} is defined by
\beq
S = - N \beta \sum_{1  \le \mu < \nu \le D}
Z_{\mu\nu} \, \Tr (U_\mu U_\nu U_\mu^\dag U_\nu^\dag )  
+ \rm{c.c.} \ ,
\label{TEKaction}
\eeq
where $U_\mu$ ($\mu = 1, \cdots , D$) are $N\times N$ unitary
matrices, and $Z_{\mu\nu} = (Z_{\nu\mu})^{*}$ is a phase
factor which is referred to as the ``twist''.
Here and henceforth we assume $D$ to be an even integer.
The model has the SU($N$) symmetry
\beq
U_\mu \mapsto g \, U_\mu \, g^\dag 
\label{SUN_gauge}
\eeq
and the U(1)$^D$ symmetry
\beq
U_\mu \mapsto \ee ^{i \alpha_\mu} U_\mu \ .
\label{U1_gauge}
\eeq
As SU($N$) covariant quantities we define
\beq
w(\mu ,\nu , \cdots , \sigma) = 
\frac{1}{N} \, 
U_\mu U_\nu  \cdots  U_\sigma   \ ,
\label{w_gauge}
\eeq
where the Greek indices may take the values
$\pm 1 \cdots \pm D$,
and we define $U_{-\mu}=U_{\mu}^\dag$.
For each term $w(\mu ,\nu , \cdots , \sigma)$, we 
introduce the displacement vector
\beq
\vec{v} = \hat{\mu} + \hat{\nu} + \cdots + \hat{\sigma} \ ,
\eeq
where $\hat{\mu}$ is a unit vector in the $\mu$ direction.
Under the U(1)$^D$ transformation (\ref{U1_gauge})
the quantity (\ref{w_gauge}) transforms as
\beq
w(\mu ,\nu , \cdots , \sigma) \mapsto
\ee^{i \vec{v} \cdot \vec{\alpha}} \, 
w(\mu ,\nu , \cdots , \sigma)  \ ,
\eeq
and $\Tr w(\mu ,\nu , \cdots , \sigma)$ is SU($N$) invariant.
The EK equivalence states that in the planar large $N$ limit
the vacuum expectation value (VEV) 
\beq
W(\mu ,\nu , \cdots , \sigma)
= \left( \prod_{\lambda \rho} (Z_{\lambda\rho})^{P_{\lambda\rho}} \right)
\langle \Tr w(\mu ,\nu , \cdots , \sigma) \rangle
\eeq
for $\vec{v}=0$
agrees with the corresponding Wilson loop in the SU($\infty$)
lattice gauge theory,
where $P_{\lambda\rho}$ denotes the number of plaquettes in the 
$(\lambda , \rho)$
plane within the surface which spans the Wilson loop.
The equivalence requires 
all the VEVs $\langle \Tr w(\mu ,\nu , \cdots , \sigma) \rangle$
for $\vec{v} \neq 0$ to vanish, which is satisfied 
if the U(1)$^D$ symmetry (\ref{U1_gauge})
is not spontaneously broken.

Now we interpret the twisted reduced model
(\ref{TEKaction}) as a NC gauge theory  \cite{AMNS}.
Note that the configuration which minimizes
the action (\ref{TEKaction}) is given by $U_\mu = \Gamma_\mu$,
where $\Gamma_\mu$ are SU($N$) matrices satisfying
the 't Hooft-Weyl algebra 
\footnote{Explicit forms of the twist $Z_{\mu\nu}$ and the corresponding
solution $\Gamma_\mu$ are given, for instance, in Refs.\ \cite{AMNS}.}
\beq
\Gamma_\mu \Gamma_\nu = Z_{\nu\mu} \Gamma_\nu \Gamma_\mu \ .
\eeq
This motivates us to make a substitution
\beq
U_\mu = \tilde{U}_\mu \Gamma_\mu  \ ,
\eeq
which brings the action (\ref{TEKaction}) into the form
\beq
S = - N \beta \sum_{1  \le \mu < \nu \le D}
\Tr \Bigl[
\tilde{U}_\mu (\Gamma_\mu \tilde{U}_\nu \Gamma_\mu^\dag)
(\Gamma_\nu \tilde{U}_\mu \Gamma_\nu^\dag)^\dag 
\tilde{U}_\nu^\dag  \Bigr]
+ \rm{c.c.} \ .
\label{TEKaction2}
\eeq
Then we consider a one-to-one correspondence
from the $N\times N$ matrices $\tilde{U}_\mu$ to 
the lattice gauge field of rank 1 on a $L^D$ lattice,
where $N^2 = L^D$ due to the matching of the degrees of freedom.
In this correspondence, 
the matrix product becomes the so-called ``star-product''
of fields on the lattice,
and the transformation
\beq
\tilde{U}_\mu \mapsto
\Gamma_\nu \tilde{U}_\mu \Gamma_\nu^\dag 
\label{transl_gauge}
\eeq
represents a shift of the lattice gauge field in the $\nu$
direction by one lattice unit.
Thus the expression (\ref{TEKaction2}) can be regarded as
the ``plaquette action'' for the NC gauge theory on the lattice.
Note also that the SU($N$) symmetry (\ref{SUN_gauge})
of the twisted reduced model can be re-written as
\beq
\tilde{U}_\mu \mapsto  g \, \tilde{U}_\mu  \, 
( \Gamma_{\mu} \, g \, \Gamma_{\mu}^\dag )^\dag  \ ,
\label{star-gauge}
\eeq
which represents the ``gauge symmetry'' of the lattice 
NC gauge theory.

The action (\ref{TEKaction2}) is invariant under
the transformation (\ref{transl_gauge}), which represents
the translational invariance of the NC gauge theory.
The whole symmetry group generated by such transformations
is $({\rm Z}_L)^D$, whose elements are given explicitly as
\beqa
\label{U_trans} 
\tilde{U}_\mu &\mapsto&
\Gamma(\vec{x}) \, \tilde{U}_\mu \, \Gamma ^\dag (\vec{x})  \\
\Gamma(\vec{x}) &=& (\Gamma_1)^{x_1} 
(\Gamma_2)^{x_2} \cdots (\Gamma_D)^{x_D} \ .
\label{def_Gamma} 
\eeqa
Note here that $(\Gamma_\mu)^L = \id$, so the $L^D$ lattice is
periodic. In terms of original variables in the twisted EK model,
the transformation (\ref{transl_gauge}) reads
\beq
U_\mu \mapsto
Z_{\nu\mu} \, \Gamma_\nu \, U_\mu \, \Gamma_\nu^\dag \ ,
\eeq
which can be realized by combining a SU($N$) transformation
(\ref{SUN_gauge}) and a U(1)$^D$ transformation (\ref{U1_gauge}).
In other words, the U(1)$^D$ symmetry of the twisted EK model
includes the translational symmetry of the NC gauge theory
up to the ``gauge symmetry'' (\ref{star-gauge}).
Thus in the new interpretation of the twisted EK model,
the NC gauge theory at $\theta = \infty$ is equivalent to
SU($\infty$) lattice gauge theory,
{\em if} the translational invariance is not spontaneously broken.

It is known that 
the U(1)$^D$ symmetry of the twisted EK model is unbroken
in both, the weak coupling (large $\beta$) and 
the strong coupling (small $\beta$)
regime \cite{Gonzalez-Arroyo:1982hz}. 
However, there is a certain indication in $D=4$
that the symmetry is broken 
at intermediate couplings \cite{Ishikawa-Okawa},
which might be related to the perturbative instabilities
pointed out in Ref.\ \cite{Guralnik:2002ru}.
These issues should be clarified further.

\section{The case of scalar field theory}

Let us move on to the case of scalar field theory.
For concreteness we discuss the ($D+1$)-dimensional NC scalar field theory,
where non-commutativity is introduced only in $D$ directions ($D$ even),
but it is straightforward to generalize our argument to the case
with an arbitrary number of commutative directions 
(including the case with {\em no} commutative direction).
%
The corresponding twisted reduced model can be given as
\cite{Procs,BHN}
\beqa
S _{\rm TRM} &=& N \, {\rm Tr} \, \sum_{t=1}^{T} 
\, \left[ \, \frac{1}{2} \,
\sum_{\mu=1}^{D} \Big( 
\Gamma_{\mu} \, \hat \phi (t) \, \Gamma_{\mu}^{\dagger}
- \hat \phi (t) \Big)^{2}  \right. \nonumber \\
&& \left.
 + \frac{1}{2} \, \Big( \hat \phi (t +1) - \hat \phi (t) \Big)^{2}
+ \frac{m^{2}}{2} \, \hat \phi (t)^{2} + 
\frac{\lambda}{4} \, \hat \phi (t)^{4} \right] \ ,  
\label{TRM_scalar}
\eeqa
where $\hat \phi (t)$ represents a Hermitian $N \times N$ matrix living
on one of the discrete time points $t = 1 , \dots , T$.
The coupling constant $\lambda$ is assumed to be real positive,
while the mass squared $m^2$ can take any real number.  
The action (\ref{TRM_scalar}) has the (Z$_L$)$^D$ symmetry 
\beq
\hat{\phi}(t) \mapsto
\Gamma(\vec{x}) \, \hat{\phi}(t) \, \Gamma ^\dag (\vec{x})  \ ,
\label{ZN2_scalar}
\eeq
which plays a crucial role.

The EK equivalence states that,
in the large $N$ limit with fixed $m^2$ and $\lambda$,
the twisted reduced model (\ref{TRM_scalar}) is ``equivalent'' to 
the matrix field theory
\beqa
S_{\rm MFT} [ \Phi ] &=& N \, {\rm Tr} \, \sum_{t=1}^{T}
\sum_{\vec{x}}  \left[ \frac{1}{2} \sum_{\mu=1}^{D}
\Big( \Phi (\vec{x} + \hat{\mu} , t) - \Phi (\vec{x}, t) \Big)^{2}
\right.  \nonumber \\
&& \left. + \frac{1}{2} 
\Big( \Phi (\vec{x} , t +1) - \Phi (\vec{x} , t) \Big)^{2}
+ \frac{m^{2}}{2} \, \Phi (\vec{x} , t)^{2} + 
\frac{\lambda}{4} \,  \Phi (\vec{x} , t)^{4} \right] 
\label{matrixfield}
\eeqa
on the $(D+1)$-dimensional hypercubic lattice, 
where $\Phi (\vec{x}  ,t)$ represents 
a Hermitian $N \times N$ matrix living on each site.
This matrix field theory has the global SU($N$) symmetry
\beq
 \Phi (\vec{x} , t) \mapsto 
g \, \Phi (\vec{x} , t) \, g^\dag \ ,
\label{SUN_mft}
\eeq
where $g\in$ SU($N$).
Let us consider a matrix field product
\beq
w(\vec{x}_1,t_1;\cdots ;\vec{x}_k,t_k )
= \frac{1}{N} \, \Phi (\vec{x}_1 , t_1)
\, \Phi (\vec{x}_2 , t_2)
\cdots
\Phi (\vec{x}_k , t_k)  \ ,
\label{w_MFT}
\eeq
whose trace
is invariant under a global SU($N$) transformation (\ref{SUN_mft}).

In the twisted reduced model (\ref{TRM_scalar}),
we may define the corresponding quantity as
\beq
\tilde{w}(\vec{x}_1,t_1;\cdots ;\vec{x}_k,t_k )
= \frac{1}{N} 
\Bigl( \Gamma(\vec{x}_1) \hat{\phi} (t_1) \Gamma ^\dag (\vec{x}_1) \Bigr)
\cdots
\Bigl(\Gamma(\vec{x}_k)\hat{\phi} (t_k) \Gamma ^\dag (\vec{x}_k) \Bigr)
\ .
\label{w_TRM} 
\eeq
Let us denote the VEV with respect to the
twisted reduced model (\ref{TRM_scalar}) and the matrix field theory 
(\ref{matrixfield}) by 
$\langle \  \cdots \ \rangle_{\rm TRM}$
and $\langle \ \cdots \ \rangle_{\rm MFT}$, respectively.
Then the statement is that
\beq
\langle \Tr \tilde{w}(\vec{x}_1,t_1;\cdots ;\vec{x}_k,t_k )
\rangle_{\rm TRM}
=
\langle \Tr w (\vec{x}_1,t_1;\cdots ;\vec{x}_k,t_k )
\rangle_{\rm MFT}
\label{EKequiv_scalar}
\eeq 
in the large $N$ limit with fixed $m^2$ and $\lambda$,
if the (Z$_L$)$^D$ symmetry (\ref{ZN2_scalar}) is not spontaneously
broken. 
A proof based on the SD equations is presented in the Appendix.

When we interpret the twisted reduced model (\ref{TRM_scalar})
as a $(D+1)$-dimensional NC scalar field theory, 
we map the matrix configuration $\{ \hat \phi (t) \} $ 
to a one-component real scalar field $\{ \phi (\vec{x} , t) \} $  
on the $L^D \times T$ lattice, where $N^2 = L^D$.
In this correspondence, 
the (Z$_L$)$^D$ symmetry (\ref{ZN2_scalar}) is nothing but
the translational symmetry in the NC directions.
Hence we conclude that the NC scalar field theory at $\theta =\infty$
is equivalent to the large $N$ matrix field theory, provided that
the translational symmetry in the NC directions is not 
spontaneously broken.

The $(2+1)$-dimensional case ($D=2$) has been studied
by Monte Carlo simulation in Refs.\ \cite{Procs,BHN}. 
In the planar large $N$ limit with 
$m^2$ below some (negative) critical value $m_{\rm c, TRM} ^2(\lambda)$,
the scalar field $\phi (\vec{x},t)$ acquires a VEV,
\footnote{As it is always the case with spontaneous breakdown of symmetries,
the VEV should be defined carefully. We introduce an explicit
symmetry breaking term in the action, calculate the VEV in the thermodynamic
limit, and finally remove the symmetry breaking term. It is assumed that
the VEVs in this article are defined in this way whenever it is necessary.}
which depends on
the coordinates $\vec{x}$ in the NC directions.
In this ``striped phase'',
the translational symmetry is spontaneously broken
\footnote{There is a numerical evidence for broken translational
invariance in 2d NC scalar field theory 
(with no commutative ``time'' direction) as well \cite{Ambjorn:2002nj}
\cite{BHN}.
As Ref.\ \cite{Ambjorn:2002nj}
pointed out, this does not contradict the Mermin-Wagner Theorem
since the proof for that Theorem assumes properties like locality and
a regular IR behavior, which do not hold in most NC field theories.
The EK equivalence has been re-examined also
in 2d chiral models \cite{Profumo:2002cm}.},
so the equivalence to the large $N$ matrix field theory is no longer
valid.

On the other hand, in the planar large $N$ limit
of the matrix field theory (\ref{matrixfield})
one expects a standard Ising-type phase transition
at some critical point $m^2 = m_{\rm c,MFT}^2(\lambda)$, below which
the matrix field $\Phi (\vec{x},t)$ 
acquires a uniform VEV proportional to 
the unit matrix.
From the simulation results of the twisted reduced model,
we may conclude that $m_{\rm c, TRM} ^2(\lambda) \ge m_{\rm c,MFT}^2(\lambda)$.
If we assume the opposite, i.e. 
$m_{\rm c, TRM} ^2(\lambda) < m_{\rm c,MFT}^2(\lambda)$, we
should have observed a uniformly ordered phase
for $m_{\rm c, TRM} ^2(\lambda) <  m^2 < m_{\rm c,MFT}^2(\lambda)$, 
which is not the case.
{\em A priori} there is no compelling reason to believe that
$m_{\rm c, TRM} ^2(\lambda) = m_{\rm c,MFT}^2(\lambda)$.

In Ref.\ \cite{BHN}
we have determined the effective mass $M_{\rm eff}$ 
(or the renormalized mass) squared
from the dispersion relation $E^2 = M_{\rm eff}^2 + \vec p ^ {\, 2}$
in the planar large $N$ limit
of the twisted reduced model (\ref{TRM_scalar}).
We have seen that it behaves 
linearly with respect to the bare mass squared as
\beq
M_{\rm eff}^2 \propto m^2 - m_{\rm c, TRM} ^2 (\lambda)
\label{Meff}
\eeq
to a good accuracy as we approach the 
critical point $m_{\rm c, TRM} ^2(\lambda)$ from above.
Since the correlation length may be identified as
$\xi = (M_{\rm eff})^{-1}$,
our result (\ref{Meff}) implies that
the (standard) critical exponent $\nu$, defined by
\beq
\xi \propto 
\Bigl\{ m^2 - m_{\rm c, TRM} ^2(\lambda) \Bigr\} ^{- \nu} \ ,
\eeq
agrees with the mean field value $\nu = 1/2$.
(More precisely, our data imply $\nu = 0.48(4)$.)

On the other hand, the critical exponent for the 
Ising-like phase transition in the 3d matrix field theory 
(\ref{matrixfield})
is expected to deviate from the mean field value generically.
For instance, Refs.\ \cite{Ferretti:1995zn,Nishigaki:1996ts}
obtain $\nu \simeq 0.67$ from a Wilsonian renormalization group study 
with a number of drastic simplifications.
If we rely on that result, then 
the fact that we obtain the mean field value $\nu \simeq 1/2$
suggests that the phase transition in the 
NC scalar field theory at $\theta =\infty$ 
has nothing to do with the phase transition
in the large $N$ matrix field theory.

\section{Summary and discussions}

In this letter
we pointed out that the
``equivalence'' between non-commutative field theories at $\theta =\infty$
and large $N$ matrix field theories, which is commonly believed 
based on perturbation theory, does not hold when the translational
symmetry is spontaneously broken.
Our argument is based on the new interpretation of the twisted reduced
models as the lattice NC field theories.
In particular the equivalence does not hold
in the striped phase of the NC scalar field theory.
\footnote
{This does not agree with a statement in Ref.\ \cite{Gubser:2000cd}
that the equivalence holds beyond the critical point,
which is based on analytic continuation in the
complex parameter space.}
Even in such a parameter region,
if one formally expands the two theories around the classical vacuum
with a {\em uniform} order,
one would obtain the same perturbative series. What occurs here is that
the NC theory actually chooses a different vacuum.
%
%

In the traditional context of EK equivalence, the spontaneous symmetry
breaking (SSB) is problematic
if it occurs at the points in the parameter space where
the continuum limit should be taken.
For instance, 
in the case of scalar field theory, one cannot study the continuum 
limit of the large $N$ matrix field theories by using the twisted
reduced models.
In the case of gauge theory, the SSB might occur at some intermediate
coupling, but this is harmless since the continuum limit is taken 
by extrapolating the bare coupling constant 
to zero along with $a \rightarrow 0$.

On the other hand, from the view point of NC field theories, 
the SSB of translational symmetry provides 
a possibility to construct NC field theories with finite $\theta$
which do not reduce to the corresponding large $N$ matrix field theories
in the $\theta \rightarrow \infty$ limit.
The 3d NC scalar field theory studied in Ref.\ \cite{BHN} 
provides the first explicit example of this sort.
This implies that the universality class of NC field theories can be
different from that of large $N$ matrix field theories.

\acknowledgments

It is our pleasure to thank Tomomi Ishikawa, 
Satoshi Iso, Hikaru Kawai, Yoshihisa Kitazawa and Yoshiaki Susaki 
for useful discussions.
The work of J.N.\ was supported in part by Grant-in-Aid for 
Scientific Research (No.\ 14740163) from 
the Ministry of Education, Culture, Sports, Science and Technology.

\appendix

\section{A proof of the EK equivalence for scalar field theory}


In this Appendix we show that the EK equivalence (\ref{EKequiv_scalar})
between the twisted reduced model (\ref{TRM_scalar})
and the matrix field theory (\ref{matrixfield})
holds in the large $N$ limit with fixed $m^2$ and $\lambda$,
if the (Z$_L$)$^D$ symmetry (\ref{ZN2_scalar})
is not spontaneously broken.
For that purpose, we consider 
the SD equations, which form a closed set of equations for 
the VEVs in eq.\ (\ref{EKequiv_scalar}).
The proof is analogous to the ones for other models
\cite{Eguchi:1982nm,Gonzalez-Arroyo:1982hz,Gonzalez-Arroyo:vx}.

In the matrix field theory (\ref{matrixfield}),
let us consider the quantity
\beq
\Bigl\langle
\Tr 
\{ \lambda_a \, \Phi (\vec{x}_1 , t_1)
\, \Phi (\vec{x}_2 , t_2)
\cdots
\Phi (\vec{x}_k , t_k) \}
\Bigr\rangle_{\rm MFT} \ ,
\eeq
where $\lambda_a$ is a generator of U($N$),
and apply the change of variable
\beq
\Phi (\vec{x}_1 , t_1) \mapsto 
\Phi (\vec{x}_1 , t_1) + \epsilon \, \lambda_a \ ,
\eeq 
where $\epsilon$ is an infinitesimal parameter.
Summing over the index $a=1,\cdots  , N^2$ and using the formula
\beq
\sum_{a=1}^{N^2} (\lambda_a)_{ij} \, (\lambda_a)_{kl}
= \delta_{il} \, \delta_{jk} 
\eeq
and the large $N$ factorization property, we obtain
\beqa
0 &=&
\Bigl\langle 
\Tr w(\vec{x}_2,t_2;\cdots ;\vec{x}_k,t_k )
\Bigr\rangle_{\rm MFT} \non
&~& + 
\Bigl\langle \Tr w(\vec{x}_1,t_1;\cdots ;\vec{x}_k,t_k ) \non
&~&  \quad \ \Bigl[ \sum_{\mu=1}^{D} \{ \Phi (\vec{x}_1+\hat{\mu} , t_1) 
+ \Phi (\vec{x}_1-\hat{\mu} , t_1) 
- 2 \Phi (\vec{x}_1 , t_1) \} \non
&~& \quad \quad + 
\{  \Phi (\vec{x}_1 , t_1+1) + 
\Phi (\vec{x}_1 , t_1-1) - 
2 \Phi (\vec{x}_1 , t_1) \}  \non
&~& \quad \quad - m^2 \Phi (\vec{x}_1 , t_1) 
    - \lambda \Phi (\vec{x}_1 , t_1) ^3  \Bigr] \Bigr\rangle_{\rm MFT} \non
&~& + \sum _{s = 2}^k
\delta_{\vec{x}_1 , \vec{x}_s } \delta_{t_1 t_s}
\Bigl\langle \Tr w(\vec{x}_1,t_1;\cdots ;\vec{x}_{s-1},t_{s-1} )
\Bigr \rangle_{\rm MFT}  \non
&~& \quad \times 
\Bigl\langle \Tr w(\vec{x}_{s+1},t_{s+1};\cdots ;\vec{x}_k,t_k ) 
\Bigr \rangle_{\rm MFT}  \ .
\label{SD_MFT}
\eeqa

SD equations for the twisted reduced model (\ref{TRM_scalar})
can be derived in a similar manner.
Let us consider the quantity
\beq
\Bigl\langle
\Tr 
\{ \Gamma(\vec{x}_1) \lambda_a \hat{\phi} (t_1) 
\Gamma ^\dag(\vec{x}_1) 
\cdots
\Gamma(\vec{x}_k) \hat{\phi} (t_k) 
\Gamma ^\dag(\vec{x}_k)  \}
\Bigr\rangle_{\rm TRM} \ ,
\eeq
and apply the change of variable
\beq
\hat{\phi} (t_1) \mapsto 
\hat{\phi} (t_1) + \epsilon \, \lambda_a \ .
\eeq 
Summing over the indices 
and using the large $N$ factorization property, we obtain
\beqa
0 &=&
\Bigl\langle 
\Tr \tilde{w}(\vec{x}_2,t_2;\cdots ;\vec{x}_k,t_k )
\Bigr\rangle_{\rm TRM} \non
&~& + 
\Bigl\langle \Tr \tilde{w}
(\vec{x}_1,t_1;\cdots ;\vec{x}_k,t_k ) \non
&~&  \quad \ \Bigl[ \sum_{\mu=1}^{D} \{ 
\Gamma_\mu \hat{\phi} ( t_1) \Gamma^\dag_\mu 
+ \Gamma_\mu^\dag \hat{\phi} ( t_1) \Gamma_\mu
- 2 \hat{\phi} (t_1) \} \non
&~& \quad \quad  + 
\{  \hat{\phi} (t_1+1) + 
\hat{\phi} (t_1-1) - 
2 \hat{\phi} (t_1) \}  \non
&~& \quad \quad  - m^2 \hat{\phi} (t_1) 
    - \lambda \hat{\phi} (t_1) ^3  \Bigr] \Bigr\rangle_{\rm TRM} \non
&~& + \sum _{s = 2}^k
\delta_{t_1 , t_s }
\Bigl\langle \Tr w(\vec{x}_1,t_1;\cdots ;\vec{x}_{s-1},t_{s-1} )
\Gamma (\vec{x}_s) \Gamma ^\dag (\vec{x}_1)
\Bigr \rangle_{\rm TRM}  \non
&~& 
\quad \times
\Bigl\langle \Tr w(\vec{x}_{s+1},t_{s+1};\cdots ;\vec{x}_k,t_k ) 
\Gamma (\vec{x}_1) \Gamma ^\dag (\vec{x}_s)
\Bigr \rangle_{\rm TRM}  \ .
\label{SD_TRM}
\eeqa

Comparing eqs.\ (\ref{SD_MFT}) and (\ref{SD_TRM}),
we find that the SD equations for the twisted reduced model
are identical to those for the matrix field theory 
if and only if
\beq
\Bigl\langle \Tr \tilde{w}
(\vec{x}_1,t_1;\cdots ;\vec{x}_{n},t_{n} ) 
\Gamma (\vec{x})
\Gamma ^\dag (\vec{x}_1)
\Bigr\rangle _{\rm TRM}
= 0  \quad \quad \mbox{for} \ \forall \vec{x} \neq \vec{x}_1 \ .
\eeq
Note that the operator on the left-hand-side transforms
non-trivially under the 
(Z$_L$)$^D$ transformation  (\ref{ZN2_scalar}),
which is a symmetry of the twisted reduced model.
Therefore the SD equations coincide
in the large $N$ limit with fixed $m^2$ and $\lambda$,
if the (Z$_L$)$^D$ symmetry is not spontaneously broken.
Assuming that the SD equations have a unique solution
\footnote{Although this assumption is valid generically,
there are special cases in which the solution is not 
unique \cite{Nishimura:1996pe}.}, 
we obtain eq.\ (\ref{EKequiv_scalar}).

\end{document}